\documentclass[12pt]{article}
\usepackage{graphicx}
\usepackage{amssymb}

\newsavebox{\sboxpubnumber}
\newsavebox{\sboxpubdate}
\newcommand{\pubdate}[1]{\begin{lrbox}{\sboxpubdate}{#1}\end{lrbox}}
\newcommand{\pubnumber}[1]{\begin{lrbox}{\sboxpubnumber}{\begin{tabular}{l} #1 \\
                 \usebox{\sboxpubdate}
                 \end{tabular}}
                           \end{lrbox}
                           \pubblock}
\newcommand{\Title}[1]{\begin{center} {\Large #1 } \end{center}}
\newcommand{\Author}[1]{\begin{center}{ \sc #1} \end{center}}
\newcommand{\Address}[1]{\begin{center}{ \it #1} \end{center}}

\newcommand{\pubblock}{\rightline{
            \usebox{\sboxpubnumber}}}
\newenvironment{Abstract}{\begin{quotation}  }{\end{quotation}}
\newenvironment{Presented}{\begin{quotation} \begin{center}
             PRESENTED BY VALERIO BOZZA AT\end{center}\bigskip
      \begin{center}\begin{large}}{\end{large}\end{center}
      \end{quotation}}

\begin{document}

\begin{titlepage}
\pubdate{\today}                    
\pubnumber{} 

\vfill \Title{Scalar fluctuations in dilatonic brane--worlds}
\vfill \Author{Valerio Bozza}

\Address{Dipartimento di Fisica ``E.R. Caianiello'', Universit\`a
di Salerno \\
         Via S. Allende, 84081 Baronissi (SA), Italy. \\
       INFN, Sezione di
Napoli, Gruppo Collegato di Salerno, Salerno, Italy.}

\vspace{.25cm} \Author{Maurizio Gasperini} \Address{Dipartimento
di Fisica, Universit\`a di Bari, \\ Via G. Amendola 173, 70126
Bari, Italy.\\ INFN, Sezione di Bari, Bari, Italy.}

\vspace{.25cm} \Author{Gabriele Veneziano} \Address{Theoretical
Physics Division, CERN \\ CH-1211 Geneva 23, Switzerland.
\\Laboratoire de Physique Th\'eorique, Universit\'e Paris Sud,
91405 Orsay, France.} \vfill

\begin{Abstract}
We derive and solve the full set of scalar perturbation equations
for a class of five-dimensional brane--world solutions, with a dilaton
scalar field coupled to the bulk cosmological constant and to a 3-brane.
The spectrum contains one localized massless scalar mode, to be
interpreted as an effective dilaton on the brane, inducing  long--range
scalar interactions. Two massive scalar modes yield corrections to
Newton's law at short distances, which persist even in the limit of
vanishing dilaton (namely, in the standard Randall--Sundrum
configuration).
\end{Abstract}
\vfill
\begin{Presented}
    COSMO-01 \\
    Rovaniemi, Finland, \\
    August 29 -- September 4, 2001
\end{Presented}
\vfill
\end{titlepage}
\def\thefootnote{\fnsymbol{footnote}}
\setcounter{footnote}{0}

\section{Introduction}

In the wake of Horava--Witten heterotic M-theory \cite{HW},
Randall and Sundrum (RS) have recently proposed a five dimensional
model where the orbifold compactification, supplemented by a
geometrical warp factor along the extra--dimension, is used to
implement an effective hierarchy between the Planck mass and
the electroweak scale \cite{RS1}. Thanks to the warp factor, on the
other hand, it is also possible to obtain consistent configurations with
only one brane and an infinite extra--dimension, while preserving the
long--range behaviour of the four dimensional Newton's law \cite{RS2}.
This is possible thanks to the dynamical localization of massless
gravitons on the brane and a strong suppression of all massive,
higher-dimensional modes.

As a consequence of the non-factorized structure of the metric in this
type of scenarios, the decomposition of the metric fluctuations,
performed according to the rotational $O(3)$ symmetry on the brane,
requires a generalized gauge--invariant formalism \cite{BDBL}. The
classical and quantum analysis of metric fluctuations is of primary
importance for understanding the possible localization of massless
modes on the brane, as well as the nature of the possible
short-range corrections  due to the continuum of massive modes, living
in the bulk. Until now, the study of this problem has been mainly
focused on the structure of tensor (i.e. transverse-traceless)
perturbations of the bulk geometry (see \cite{CEHS} for a general
discussion).

In all string/M-theory models, however, the graviton is
accompanied by massless scalar partners (the dilaton, the
compactification moduli, etc.). These typically induce long-range
interactions of gravitational strength \cite{TV}, not (yet)
experimentally observed. The conventional  way to solve this problem
is to assume that the scalar partners get a non-perturbative,
SUSY-breaking mass, thus suppressing the range of the associated
scalar interactions. However, if scalar fluctuations would not be
confined in RS--type brane--worlds, then all corresponding interactions
on the brane would be suppressed,  and the brane-world scenario could
naturally solve a possible discrepancy between String/M-theories and
experiments.

Here we present the results of Ref. \cite{BGV}, where we discuss
the localization of scalar fluctuations in a typical
brane-world scenario of the RS type, taking into account the
possible existence of scalar sources and scalar fields living in
the bulk. To this purpose, we shall consider a non-compact,
$\mathbb{Z}_2$-symmetric, five-dimensional background, generated
by a positive tension 3-brane and by a bulk dilaton field coupled
to the brane and to the (negative) bulk energy density. We shall
restrict ourselves to the gravi-dilaton solutions discussed in
\cite{CLP}, which generalize the AdS$_5$ RS scenario in the
presence of a bulk scalar field, and which are already known to
guarantee the localization of tensor metric fluctuations.

\section{Background equations}

We consider in particular a five-dimensional scalar-tensor background
$\{g_{AB}, \phi\}$, possibly arising from the bosonic sector of a
dimensionally reduced string/supergravity theory, and
non-trivially coupled to a negative cosmological constant
$\Lambda$ and to a 3-brane of  positive tension $T_3$:
\begin{eqnarray}
 S=S_{\rm bulk}+
S_{\rm brane}&=&{M_5^3}\int d^5 x \sqrt{|g|} \left( -R+\frac{1}{2}
g^{AB}
\partial_A \phi
\partial_B \phi -2 \Lambda e^{\alpha_1 \phi}
\right) \nonumber\\ &-& \frac{T_3}{2} \int d ^4 \xi
\sqrt{|\gamma|}\left[ \gamma^{\alpha \beta} \partial_\alpha X^A
\partial_\beta
 X^B g_{AB}
e^{\alpha_2 \phi} -2 \right]. \label{21}
\end{eqnarray}
Here $M_5$ is the fundamental mass scale of the five-dimensional
bulk space-time, and the parameters $\alpha_1, \alpha_2$ control
the coupling of the bulk dilaton to $\Lambda$ and to the brane. We
allow in general for non-minimal couplings (a single exponential
potential for the dilaton can also be derived from the dimensional
reduction of a suitable higher-dimensional model \cite{CLP}). The
brane action is parameterized by the coordinates $X^A(\xi)$
describing the embedding of the brane in the bulk manifold, and by
the auxiliary metric tensor $\gamma_{\alpha \beta}(\xi)$ defined
on the four-dimensional world-volume of the brane, spanned by the
coordinates $\xi^\alpha$.

In our conventions, greek indices run from $0$ to $3$, capital
Latin indices from $0$ to $4$, lower-case Latin indices from $1$
to $3$. For the bulk coordinates we use the notation $x^A=(t,x^i,z)$.

The field equations are obtained by the variation of the action
with respect to $g_{AB}$, $\phi$, $X^A$ and
$\gamma_{\alpha\beta}$, respectively. These equations can be
specialized to the case of a conformally flat background, with
warp factor $a(z)$ and a dilaton $\phi(z)$. Also, we shall look
for $\mathbb{Z}_2$--symmetric solutions, describing a flat brane
rigidly located at $z=0$, and we set
\begin{equation}
g_{AB}= a^2(z)\eta_{AB}, ~~~~~~ \phi = \phi(z), ~~~~~~ X^A =
\delta^A_\mu\xi^\mu. \label{26}
\end{equation}
where $\eta_{AB}$ is the five-dimensional Minkowski metric. The
induced metric thus reduces to
\begin{equation}
\gamma_{\alpha \beta} = \delta^A_\alpha \delta^B_\beta g_{AB}~
e^{\alpha_2 \phi},
\end{equation}
while the brane equations are identically satisfied thanks to the
$\mathbb{Z}_2$ symmetry.

The dynamical equations are obtained from the dilaton equation,
which becomes
\begin{equation}
3\frac{a'}{a}\phi'+\phi''-2\alpha_1 \Lambda a^2 e^{\alpha_1
\phi}-2\alpha_2 T_3 a e^{2\alpha_2 \phi} \delta(z)=0 , \label{Eq
reduced dilaton}
\end{equation}
and from the $(\alpha,\beta)$ and $(4,4)$ components of the
Einstein equations, which give, respectively,
\begin{eqnarray}
 && -3\frac{a''}{a} =\frac{\phi'^2}{4}+\Lambda a^2
e^{\alpha_1 \phi}+\frac{T_3}{2}a e^{2\alpha_2 \phi} \delta(z),
\label{Eq G00}\\ && -6\frac{a'^2}{a^2} =
-\frac{\phi'^2}{4}+\Lambda a^2 e^{\alpha_1 \phi}
 \label{Eq G44}
\end{eqnarray}
(a prime denotes differentiation with respect to $z$).

If we fine-tune the parameters by choosing
\begin{equation}
\alpha_1= 4 \alpha_2 \;,~~~ T_3 = 8 \sqrt{\Lambda/\Delta}\;,~~~
\alpha_1^2=\Delta+\frac{8}{3}, \label{alpha rel}
\end{equation}
where the last equation defines $\Delta$, we recover a
one-parameter family of exact domain wall solutions (hereafter,
CLP solutions) \cite{CLP}. The solution corresponds to a brane of
positive tension, $T_3>0$, provided $\Delta \leq -2$. This range
of $\Delta$ guarantees a positive tension and also avoids the
presence of naked singularities \cite{CLP}. On the other hand,
when $\alpha_1$ goes to zero, i.e. when $\Delta = -8/3$, the
dilaton decouples and the background reduces to Randall-Sundrum
one \cite{RS2}. We shall thus assume${-8\over 3} \leq \Delta \leq
-2$.

\section{Scalar perturbations}

We perturb to first order the full set of bulk equations, keeping
the position of the brane fixed, $\delta X^A=0$. We thus set
\begin{equation}
\delta g_{AB}=h_{AB}, ~~~~~~ \delta g^{AB}=-h^{AB}, ~~~~~~
\delta\phi=\chi , ~~~~~~ \delta X^A=0, \label{Pert dil}
\end{equation}
where the indices of the perturbed fields are raised and lowered
by the unperturbed metric, and the background fluctuations
$h_{AB}, \chi$ are assumed to be inhomogeneous.

We shall expand around the CLP background  in the so-called
``generalized longitudinal gauge" \cite{BDBL}, which extends the
longitudinal gauge of standard cosmology \cite{MBF} to the
brane-world scenario. As discussed in \cite{BDBL}, in five
dimensions there are four independent degrees of freedom for the
scalar metric fluctuations: in the generalized longitudinal gauge
they are described by the four variables $\{ \varphi, \psi,
\Gamma, W\}$, defined by
\begin{eqnarray}
&& h_{00}= 2 \varphi a^2, ~~~~~~~~~~~~~~~ h_{ij}=2\psi a^2
\delta_{ij}, \nonumber\\ && h_{44}= 2 \Gamma a^2, ~~~~~~~~~~~~~~~
h_{04}=-W a^2. \label{38}
\end{eqnarray}
The perturbation of the backgraound equations leads then to the full set
of constraints and dynamical equations governing the linearized
evolution of the five scalar variables $\{ \varphi, \psi, \Gamma,
W,\chi\}$.

In the absence of bulk sources with anisotropic stresses we
can eliminate $\varphi$ from Einstein the eq. ($i \not=j$), thus reducing
the system to four scalar degrees of freedom by setting:
\begin{equation}
\varphi=\psi+\Gamma. \label{318}
\end{equation}
As a consequence, we find that the variable $W$
decouples from the other fluctuations:
\begin{equation}
\square_5 W=3\left( \frac{a''}{a}-\frac{a'^2}{a^2} \right)W
\label{319}
\end{equation}
but, because of the non-trivial self-interactions, $W$ does not freely
propagate in the background geometry like the graviton, which
statisfies the pure  five-dimensional D'Alembert equation
\begin{equation}
\square_5 h_{ij}=0 \label{Eq libera},
\end{equation}
 where $\square_5  = \nabla_M \nabla^M$.

In order to discuss the dynamics of the remaining variables $\psi,
\Gamma$ and $\chi$, it is now convenient to recombine their
differential equations in an explicitly covariant way,  to obtain
canonical evolution equation. We find :
\begin{equation}
\square_5 \psi = f_{\psi}(\Gamma,\chi), ~~~~~~ \square_5 \Gamma =
f_{\Gamma}(\Gamma,\chi), ~~~~~~ \square_5 \chi =
f_{\chi}(\Gamma,\chi),
\label{Eq psiGammachi}
\end{equation}
where the source terms
 depend only on $\Gamma$ and $\chi$. By introducing the fields
\begin{equation}
\omega_1=2\psi+ \Gamma, ~~~~~ \omega_2=6\alpha_2 \Gamma +\chi,
~~~~~ \omega_3=\Gamma- 2\alpha_2 \chi, \label{323}
\end{equation}
the above system of coupled equations can be diagonalized, and the
perturbation equations (\ref{Eq psiGammachi}) reduce to
\begin{equation}
\square_5 \omega_1=0, ~~~~~~ \square_5 \omega_2=0 , ~~~~~~
\square_5 \omega_3= f_\omega (\omega_3).  \label{Eq omega}
\end{equation}

Together with eq. (\ref{319}), and the constraints between these
fields coming from the Einstein equations, such decoupled
equations describe the complete evolution of the scalar (metric +
dilaton) fluctuations in the CLP brane-world background. Two
variables ($\omega_1, \omega_2$) are (covariantly) free on the
background like the graviton, while the other two variables
($\omega_3, W$) have non-trivial self-interactions.

In all cases, it is convenient to introduce the corresponding ``canonical
variables" $\hat W$, $\hat \omega_i$ ($i=1,2,3$), which have
canonically normalized   kinetic terms \cite{MBF} in the action,
simply by absorbing the geometric warp factor as follows:
\begin{equation}
W= \hat W a^{-3/2}, ~~~~~~~~~~~~~~~ \omega_i= \hat \omega_i
a^{-3/2}. \label{328}
\end{equation}
When the general solution is written as a  superposition of free,
factorized plane-waves modes on the brane,
\begin{equation}
\hat W= \Psi_w(z)  e^{-ip_\mu x^\mu}, ~~~~~~~~~~~ \hat \omega_i=
\Psi_i(z) e^{-ip_\mu x^\mu}
\label{329}
\end{equation}
they define the inner product of states with measure $dz$, as in
conventional one-dimensional quantum mechanics. Such a product is
required for an appropriate definition of normalizable solutions.

The allowed mass spectrum of $m^2=\eta^{\mu\nu} p_\mu p_\nu$, for
the scalar fluctuations on the brane, can then be obtained by
solving an eigenvalue problem in the Hilbert space $L^2(R)$ for
the canonical variables $\Psi_w,\Psi_i$, satisfying a
Schr\"odinger-like equation in $z$, which is obtained from the
equations (\ref{319}), (\ref{Eq omega})  for $W$ and $\omega_i$,
and which can be written in the conventional form as:
\begin{equation}
\Psi_w''+\left(m^2- {\xi_w''\over \xi_w}\right)\Psi_w=0,
~~~~~~~~~~~~ \Psi_i''+\left(m^2- {\xi_i''\over
\xi_i}\right)\Psi_i=0. \label{330}
\end{equation}
Here, by analogy with cosmological perturbation theory \cite{MBF},
we have introduced four ``pump fields" $\xi_w,\xi_i$, defined  as
follows:\begin{eqnarray} & \xi_w=a^{\beta_w}, ~~~~~~~~~~~~
\xi_i=a^{\beta_i},& \nonumber\\ & \beta_w=-{3\over 2},~~~
\beta_1=\beta_2 ={3\over 2}, ~~~ \beta_3=-{1\over 2}(1+3
\alpha_1^2)=-{3\over 2}(\Delta+3). &\label{332} \end{eqnarray}

The effective potential generated by the derivatives of the pump
fields depends on $\beta_w,\beta_i$, and contains in general a
smooth part, peaked at $z=0$, plus a positive or negative
$\delta$-function contribution at the origin. We may have, in
principle, not only volcano-like potentials, which correspond to
the free covariant d'Alembert equation with $\beta=3/2$ \cite{CLP}
(and which are known to localize gravity \cite{RS2,CEHS}), but
also potentials that are positive everywhere and admit no bound
states.

\section{Localization of the massless modes}

The general solutions of the canonical perturbation equations
(\ref{330}) are labelled by the mass eigenvalue $m$, by their
parity with respect to $z$-reflections, and by the parameters
$\beta_w,\beta_i$, which depend on the type of perturbation. To
describe a bound state, we shall restrict such solutions to those
with a normalizable canonical variable with respect to the measure
$dz$, namely to $\Psi(z) \in L^2(R)$. Among the acceptable
solutions, we shall finally select those satisfying all the
constraints coming from Einstein equations. The above set of
conditions will determine the class of brane-world backgrounds
allowing the four-dimensional localization of long-range scalar
interactions.

Analyzing Eqs. (\ref{330}) for $m=0$, we find that $W$ and $\omega_3$
are not normalizable. The even solutions of the free d'Alembert equation
are instead normalizable, so we have acceptable solutions  for
$\omega_1$ and $\omega_2$. However, because of the constraints,
$\omega_1$ is forced to vanish when $W=0$, unless the fluctuations
are static, $\dot \omega_1=0$. It follows that there are two
independent massless modes localized on the brane: one,
$\omega_2$, is propagating and the other, $\omega_1$,  is static.

We may thus conclude that all CLP backgrounds with $-8/3 \leq
\Delta \leq -2$ localize on the brane not only the massless spin-2
degrees of freedom \cite{CLP}, but also one propagating massless
scalar degree of freedom ($\omega_2$), corresponding to a long-range
scalar interaction generated by the dilaton field. The second
independent massless degree of freedom localized on the brane
($\omega_1$) is not propagating ($\dot \omega_1=0$), but is essential
to reproduce the standard long-range gravitational interaction in the
static limit, as we shall discuss later. In the limiting case of a pure
AdS$_5$ solution  ($\Delta=-8/3$) the dilaton disappears from the
background, and the dilaton fluctuation $\chi=\omega_2$ decouples
from the others. The only (static) contribution to the scalar
sector of metric fluctuations comes from $\omega_1$, which
generates the long-range Newton potential $\varphi=\psi$ on the
brane.

\section{The massive mode spectrum}

The massive part of the spectrum of the canonical equations
(\ref{330}) is not localized on the brane; it may induce
higher-dimensional, short-range corrections to the long-range
scalar forces, which represent physical effects from
the fifth dimension on our brane.

It is important to note that modes with negative squared mass
(tachyons) are not included in the spectrum, as they would not
correspond to a normalizable canonical variable ($\Psi$ would blow
up in $z$). Another consequence of the normalization condition is
the mass gap between the localized massless mode and the massive
corrections, in the limiting background with $\Delta=-2$ (already
noticed in \cite{CLP} for the case of pure tensor interactions).

Imposing all the constraints from Einstein equations, we find
that, in contrast with the massless case, none of the four scalar
fluctuations is forced to vanish. However, only two amplitudes are
independent. By taking, for instance, $\omega_2$ and $\omega_3$ as
independent variables, we can indeed express $W$ and $\omega_1$ in
terms of the other fields, for all values of $\Delta$.

For such backgrounds we  thus have four types of
higher-dimensional contributions to the scalar interactions on the
brane, arising from the massive spectrum of $\omega_i$ and $W$.
The massive $\omega_2$ modes can be considered as the KK
excitations of the effective dilaton zero--mode localized on the
brane. The massive $\omega_3$ are still present even if we turn
off the dilaton and are therefore a product of the
extra--dimension itself. In the limiting RS case, $\omega_3\equiv
\Gamma$ is the breathing mode of the fifth dimension, which
induces short range corrections to the gravitational force.

\section{Static limit and leading-order corrections}

We can now compute the effective scalar-tensor interaction
induced on the brane, in the weak field limit, by a static and
point-like source of mass $M$ and dilatonic charge $Q$.

In our longitudinal gauge (\ref{38}), in which the decomposition
of the metric fluctuations is based on the $O(3)$ symmetry of the
spatial hypersurfaces of the brane, the energy density of a
point-like particle only contributes to the scalar part of the
perturbed matter stress tensor (with $T_{00}$ as the only
non-vanishing component), and provides a $\delta$-function source
to the $(0,0)$ scalar perturbation equation. Similarly, the charge
$Q$ acts as a point-like source in the dilaton perturbation
equation.

As a  consequence, we obtain three $\delta$-function sources in
the equations for the three $\omega_i$ fluctuations, $S_i
\delta^3(x-x')\delta(z)$, with three scalar charges $S_i$, which
are ``mixtures"  of $M$ and $Q$, while no source term is obtained
in the static limit for the $W$ fluctuation.

The exact static solutions of eqs. (\ref{Eq omega}) can be easily
obtained using the static limit of the retarded Green function
evaluated on the brane ($z=0$), i.e.
\begin{equation} \omega_i(x,x')= -S_i G_i (\nu,x,x'), \label{62}
\end{equation}
where
\begin{equation}
G_i (\nu,\vec x, \vec x', z=z'=0)= \int {d^3p\over (2 \pi)^3} e^{i
\vec p \cdot (\vec x -\vec
x')}\left\{{\left[\psi^+_{0}(0)\right]^2\over p^2}
+\int_{m_0}^\infty dm {\left[\psi^+_{m}(0)\right]^2\over p^2+m^2}
\right\},
 \label{63}
\end{equation}
is the static Green function, constructed by the exact massless
and massive eigenfunctions $\psi^+_{0}(z), \psi^+_{m}(z)$ of Eqs.
(\ref{330}) (respectively). The first term in the integrand
corresponds to the long-range forces generated by the massless
modes, the second term to the ``short-range" corrections due to
the massive modes, and $m_0$ is the lower bound for the massive
spectrum ($m_0=k/2$ if $\Delta=-2$, while $m_0=0$ if $\Delta \neq
-2 $).

We should note that in the $\omega_1, \omega_2$ case we have to
include both the massless and massive contributions, while in the
$\omega_3$ case only the massive ones survive. Here we report the
$\omega_i$ solutions in the case $\Delta<-2$:
\begin{eqnarray} && \omega_1=-{S_1 A_{\nu_0}\over r} \left[1+
B_{\nu_0}\left(1\over kr\right)^{2\nu_0-2}\right], \nonumber\\ &&
\omega_2=-{S_2 A_{\nu_0}\over r} \left[1+ B_{\nu_0}\left(1\over
kr\right)^{2\nu_0-2}\right], \nonumber\\ && \omega_3=-{S_3
A_{\nu_0}\over r} B_{\nu_0}\left(1\over kr\right)^{2\nu_0-2} ,
\label{610}
\end{eqnarray}
where $\nu_0=\frac{\Delta}{2(\Delta+2)}$ and the constants
$A_{\nu_0}$ and $B_{\nu_0}$ depend, through
$\nu_0$,  on the particular values of the dilaton coupling parameters,
and are fixed by the correct normalization of the canonical ``wave
function" \cite{BGV}.

Introducing explicitly the four-dimensional gravitational constant
$G$, and going back through the transformation (\ref{323}), the
scalar and dilaton fluctuations can finally be written in the form
\begin{eqnarray} && \varphi =-{GM\over r}\left[1+ {2 \alpha_2\over
1+12 \alpha_2^2} \left({Q\over M} + 2 \alpha_2\right) + {4\over
3}B_{\nu_0} \left(1\over kr\right)^{2\nu_0-2}\right], \nonumber\\
&& \psi =-{GM\over r}\left[1-{2 \alpha_2\over 1+12 \alpha_2^2}
\left({Q\over M} + 2 \alpha_2\right) + {2\over 3}B_{\nu_0}
\left(1\over kr\right)^{2\nu_0-2}\right], \nonumber\\ && \Gamma
=-{GM\over r}\left[{4 \alpha_2\over 1+12 \alpha_2^2} \left({Q\over
M} + 2 \alpha_2\right) + {2\over 3}B_{\nu_0} \left(1\over
kr\right)^{2\nu_0-2}\right], \nonumber\\ && \chi =-{GQ\over
r}\left[{2 \over 1+12 \alpha_2^2} \left(1+ 2\alpha_2 {M\over
Q}\right) + {2}B_{\nu_0} \left(1\over kr\right)^{2\nu_0-2}\right],
\label{616}
\end{eqnarray}
where $G$ is related to
the five-dimensional mass scale $M_5$ through the appropriate
integral over the ``warped" volume external to the brane.  It should be
noted that the short-range corrections induced by the massive scalar
modes have the same qualitative behaviour as in the tensor case,
discussed in \cite{CLP}, in spite of the fact that the massive scalar
modes have different spectra.

The limiting case $\Delta=-8/3$ corresponds to a pure AdS$_5$
background, if there are no scalar charges on the brane. In that
case $\omega_2$ exactly corresponds to the dilaton fluctuation
$\chi$ (see eq. (\ref{323})), and can be consistently set to zero
(toghether with the dilaton background) if we want to match, in
particular, the ``standard" brane-world configuration originally
considered by Randall and Sundrum \cite{RS2}. In this limit,
$B_{2}=1/2$, and we exactly recover previous results for the
effective gravitational interaction on the brane \cite{27a}, i.e.
\begin{equation} \varphi= -{GM\over r} \left(1+ {2\over 3
k^2r^2}\right), ~~~~~~ \psi= -{GM\over r} \left(1+ {1\over 3
k^2r^2}\right). \label{617}
\end{equation}

The massless-mode  truncation reproduces in this case the static,
weak field limit of linearized  general relativity. The massive
tower of scalar fluctuations, however, induces deviations from
Einstein gravity already in the static limit (as noted in
\cite{27a}), and is the source of a short-range force due to the
``breathing" of the fifth dimension,
\begin{equation} \Gamma= - {GM\over 3 r}{1\over (kr)^2}, \label{618} \end{equation} even
in the absence of bulk scalar fields, and of scalar charges  for
the matter  on the brane.

In a more general gravi-dilaton background ($\Delta \not= -8/3$),
the static expansion (\ref{616}) describes an effective
scalar-tensor interaction on the brane, which is potentially
dangerous for the brane-world scenario, as it contains not only
short-range corrections, but also long-range scalar deviations
from general relativity (and, possibly, violations of the Einstein
equivalence principle), even in the interaction of ordinary
masses, i.e. for $Q=0$. This seems to offer an interesting window
to investigate the effects of the bulk geometry on the
four-dimensional  physics of the brane.

\section{Conclusions}

We have analyzed the full set of coupled equations governing the
evolution of scalar fluctuations in a dilatonic brane-world
background, supporting a flat 3-brane rigidly located at the fixed
point of $Z_2$ symmetry. We have diagonalized the system of
dynamical equations, and found four decoupled but self-interacting
variables representing, in a five-dimensional bulk, the four
independent degrees of freedom of scalar excitations of the
gravi-dilaton background.

We have presented the exact solutions of the canonical
perturbation equation for all the scalar degrees of freedom, and
we have discussed, the effects of their massless and massive
spectrum for the scalar interactions on the brane.

For all dilaton couplings, there is  one propagating massless mode
localized on the brane,  associated with a long-range dilatonic
interaction in four dimensions. In addition, KK modes in the bulk
yield short--range scalar corrections to the Newton's law. These
corrections are present even in the RS limit where the dilaton
vanishes, because of the breathing mode of the fifth dimension.

Finally, we note that our results are different from those
obtained in the case of thick branes with a confining scalar
potential \cite{Giovannini}.

\end{document}